# Habitability Models for Planetary Sciences

Abel Méndez, Planetary Habitability Laboratory, University of Puerto Rico at Arecibo, USA (abel.mendez@upr.edu)
Edgard G. Rivera-Valentín, Lunar and Planetary Institute, USRA, USA
Dirk Schulze-Makuch, Technical University Berlin, Germany
Justin Filiberto, Lunar and Planetary Institute, USRA, USA
Ramses Ramírez, Earth-Life Science Institute, Japan
Tana E. Wood, International Institute of Tropical Forestry, USDA Forest Service, Puerto Rico, USA
Alfonso Dávila, NASA Ames Research Center, USA
Chris McKay, NASA Ames Research Center, USA
Kevin Ortiz Ceballos, Planetary Habitability Laboratory, University of Puerto Rico at Arecibo, Puerto Rico, USA
Marcos Jusino-Maldonado, Planetary Habitability Laboratory, University of Puerto Rico at Arecibo, Puerto Rico, USA
Guillermo Nery, University of Puerto Rico at Arecibo, Puerto Rico, USA
René Heller, Max Planck Institute for Solar System Research, Germany
Paul Byrne, North Carolina State University, USA
Michael J. Malaska, Jet Propulsion Laboratory / California Institute of Technology, USA
Erica Nathan, Brown University, USA
Marta Filipa Simões, State Key Laboratory of Lunar and Planetary Sciences, China
André Antunes, State Key Laboratory of Lunar and Planetary Sciences, China
Jesús Martínez-Frías, Instituto de Geociencias (CSIC-UCM), Spain
Ludmila Carone, Max Planck Institute for Astronomy, Germany
Noam R. Izenberg, Johns Hopkins Applied Physics Laboratory, USA
Dimitra Atri, Center for Space Science, New York University Abu Dhabi, United Arab Emirates
Humberto Itic Carvajal Chitty, Universidad Simón Bolívar, Venezuela
Priscilla Nowajewski-Barra, Fundación Ciencias Planetarias, Chile
Frances Rivera-Hernández, Dartmouth College, USA
Corine Brown, Macquarie University, Australia
Kennda Lynch, Lunar and Planetary Institute, USRA, USA
David Catling, University of Washington, USA
Jorge I. Zuluaga, Institute of Physics / FCEN - Universidad de Antioquia, Colombia
Juan F. Salazar, GIGA, Escuela Ambiental, Facultad de Ingeniería, Universidad de Antioquia, Colombia
Howard Chen, Northwestern University, USA
Grizelle González, International Institute of Tropical Forestry, USDA Forest Service, Puerto Rico, USA
Madhu Kashyap Jagadeesh, Jyoti Nivas College, Bengaluru, India
Rory Barnes, University of Washington, USA
Charles S. Cockell, UK Centre for Astrobiology, UK
Jacob Haqq-Misra, Blue Marble Space Institute of Science, USA

**Abstract:** Habitability has been generally defined as the capability of an environment to support life. Ecologists have been using Habitat Suitability Models (HSMs) for more than four decades to study the habitability of Earth from local to global scales. Astrobiologists have been proposing different habitability models for some time, with little integration and consistency between them and different in function to those used by ecologists. In this white paper, we suggest a mass-energy habitability model as an example of how to adapt and expand the models used by ecologists to the astrobiology field. We propose to implement these models into a NASA Habitability Standard (NHS) to standardize the habitability objectives of planetary missions. These standards will help to compare and characterize potentially habitable environments, prioritize target selections, and study correlations between habitability and biosignatures. Habitability models are the foundation of planetary habitability science. The synergy between the methods used by ecologists and astrobiologists will help to integrate and expand our understanding of the habitability of Earth, the Solar System, and exoplanets.

# 1. Introduction

Life on Earth is not equally distributed. There is a measurable gradient in the abundance and diversity of life from deserts to rain forests (*i.e.*, spatially), and temporally among seasons and, at large time-scales, in geological time. Our planet also has experienced global environmental changes from the Archean to the Anthropocene. In general, *a habitable environment is a spatial region that might support some form of life* (Farmer, 2018), *albeit not necessarily one with life*. One of the biggest problems in astrobiology is how to define and measure the habitability not only of terrestrial environments but also of planetary environments, from the Solar System to extrasolar planets. The word *Habitability* literally means the *quality of habitat* (the suffix *-ity* means quality, state, or condition). Astrobiologists have been constructing different general definitions of habitability, not necessarily consistent with one another, for some time (*e.g.*, Shock & Holland, 2007; Hoehler, 2007; Cardenas *et al*., 2014; Cockell *et al*., 2016; Cárdenas *et al*., 2019; Heller, 2020). Other more specific habitability definitions, such as the canonical Habitable Zone (*i.e.*, presence of surface liquid water on Earth-like planets), are used in exoplanet science (Kasting et al., 1993). Ecologists developed a standardized system for defining and measuring habitability in the early 1980s; however, this is seldom utilized in the astrobiology community (USFWS, 1980).

The popular term habitability is formally known as *habitat suitability* in biology. Ecologists before the 1980s were using different and conflicting measures of habitability, a situation not much different than today for astrobiologists. The U.S. Fish and Wildlife Service (USFWS) decided to solve this problem with the development of the Habitat Evaluation Procedures (HEP) standards in 1974 for use in impact assessment and project planning (USFWS, 1980). These procedures include the development and application of Habitat Suitability Models (HSM) (Hirzel & Lay, 2008). Other names for these models are Ecological Niche Models (ENM), Species Distribution Models (SDMs), Habitat Distribution Models (HDM), Climate Envelope Models (CEM), Resource Selection Functions (RSF), and many other minor variants (Guisan *et al*., 2017). These multivariate statistical models are widely used today by ecologists to quantify species-environment relationships from the ground to satellite observations. Habitat Suitability Models integrate concepts as needed from ecophysiology, niche theory, population dynamics, macroecology, biogeography, and the metabolic theory of ecology.

Astrobiologists have largely not utilized HSMs for at least three reasons. First is the naming: habitability is a common word in Earth and Planetary Science, but it is not generally used by biologists. Thus, a quick review of the scientific literature shows no definition of this concept in biological terms. The second reason is the specialization: HSM is a specialized topic of theoretical ecology, which is not highly represented in the astrobiology community. The third is applicability: HSMs are mostly used to study the distribution of wild animals and plants, not microbial communities (generally the focus of astrobiological studies), so it may not seem readily applicable to the field of astrobiology. Yet endosymbiotic relationships between microorganisms (bacteria, fungi and other unicellular life) with animals and plants also play a key role in the survival of the latter. Thus, anything that can be said about habitability at the macroscopic level is tightly coupled to habitability at the microscopic level. In one way, the mathematical framework behind HSM is easier to apply to microbial communities than animals because the spatial interactions of animals (*e.g.*, predation) tend to be much more complex. However, microbial life is not easy to quantify in free-living populations and it is thus harder to validate the HSMs with them, although molecular methods are changing this (Douglas, 2018).

The definition and core framework of HSMs can be extended from the Earth to other planetary environments. However, the astrobiology field does not have the luxury of validating HSMs with the presence of life unless when applied to environments on Earth (*e.g.*, extreme environments). Thus, known ecophysiology models are used instead to predict the occurrence, distribution, and abundance of putative life in any planetary environment. A common assertion is that it is not possible to measure the habitability of a system without knowing all the environmental factors controlling it. However, even in scenarios for Earth, the approach is to select a minimum set of relevant factors to simplify the characterization of the systems. While the objective can be to establish if a system is habitable, it can, alternately, be simply to explore how the selected environmental variables contribute to the habitability of the system. Usually, a library of habitability metrics is created for each environment or lifeform under consideration, with each metric depending on the species, the scales, or the environmental factors under consideration. In a fundamental sense, the only way to really know if a place is habitable or not is to find (or put) life on it (Zuluaga *et al*., 2014; Chopra & Lineweaver, 2016). It is nearly impossible, nor is it desirable, to include all factors affecting habitability in a model, even for environments on Earth. Thus, the objective of habitability models is to understand the contributions of a *finite* set of variables toward the *potential* to support a specific species or community (e.g., primary producers, organisms that use abiotic sources of energy) (Guisan *et al*., 2017). So, even if we do not know or include all the relevant factors, we can consider the effects of those we do know.

Here we recommend adapting and expanding the ecologists' nearly four decades of experience modeling habitability on Earth to astrobiological studies. These models can be used to characterize the spatial and temporal distribution of habitable environments, identify regions of interest in the search for life, and, eventually, explore correlations between habitability and biosignatures. For example, such models would help to test the hypothesis that biosignatures (or *biomarkers*) are positively correlated with proxy indicators of geologically habitable environments (or *geomarkers*); *i.e.,* there is life whenever there are habitable environments on Earth (Martinez-Frias *et al*., 2007). Measurements by past and future planetary missions can be combined into a standard library of habitability models. Results from different missions can then be compared, even using different measurements, since, through the use of HSMs, their results can be mapped to the same standard scale (*e.g.*, zero for worst and one for best regions). A Habitability Readiness Analysis (HRA) of any mission could be used to determine how its existing instruments could be used, or what sensors should be added, for measurements in the spatial and temporal habitability scales of interest. Furthermore, it might also be possible to develop new sensors for direct habitability measurements.

This white paper to the Planetary Science and Astrobiology Decadal Survey 2023–2032 addresses many of the misconceptions about habitability and attempts to create a standard conceptual framework to assess habitability issues for future purposes. This contribution is relevant for roving, landed, or orbital missions for any planetary target of astrobiological interest, including environments on Earth. Section 2 gives an overview of current ecology models and Section 3 presents some examples of how these models are currently implemented in the astrobiology field. Section 4 describes our proposed general mass-energy approach to model habitability. Section 5 presents our specific recommendations to the Decadal Survey. Section 6 proposes science questions to be addressed by these models, beyond those traditionally associated with habitability studies. Finally, Section 7 presents our concluding remarks.

## 2. Habitability in Biology: The Habitat Suitability Models

Habitat Suitability Models (HSMs) are widely used in ecology to study the habitability of environments, many times under different definitions: species distribution models (SDMs) or environmental niche models (ENMs) (Kuhn *et al.*, 2016; Guisan *et al.*, 2017). An important step in the construction of HSMs is the selection of spatially explicit environmental variables at the right resolution to determine a species' preferred environments (*i.e.*, its niche) as close to its ecophysiological requirements as possible. Environmental variables (such as edaphic factors in soils) can exert complex direct or indirect effects on species (*e.g.*, Oren, 1999; Oren, 2001, Rajakaruna & Boyd, 2008). These variables are ideally chosen to reflect the three main types of influence on a species: (1) regulators or limiting factors, defined as factors controlling a species' metabolism (*e.g.*, temperature); (2) disturbances, defined as all types of perturbations affecting environmental systems; and (3) resources, defined as all compounds that can be consumed by organisms (*e.g.*, nutrients). There are many other variables that exert an indirect, rather than a direct, effect on species distribution. The construction of HSMs follows five general steps: (1) conceptualization; (2) data preparation; (3) model calibration; (4) model evaluation; and (5) spatial predictions (Guisan *et al*., 2017).

One of the main HSM tools is the *Habitat Suitability Index* (HSI), which provides one way to quantify the capacity of a given habitat to support a selected species. An index is the ratio of a value of interest divided by a standard of comparison. The value of interest is an estimate or measure of the quality of habitat conditions for a species in the studied environment, and the standard of comparison is the corresponding value for the optimum habitat conditions for the same evaluated species. An HSI of zero (minimum value) represents a totally unsuitable habitat, and a maximum value of one represents an optimum habitat. In developing an HSI we should obtain a direct and linear relationship between the HSI value and the carrying capacity of the environment for the species under consideration (USFWS, 1980). The functions describing the species distribution or abundance along each environmental variable in an HSM are called *species response curves* (Austin & Gaywood, 1994). These curves, when plotted, can vary from simple box-like envelopes resulting in binary indices to more gradual and complex responses resulting in continuous indices.

Carrying capacity is generally defined as the maximum supported population density in equilibrium. More precisely, carrying capacity is the user-specified quality biomass of a particular species for which a particular area will supply all energetic and physiological requirements over a long, but specified, period (Giles, 1978). Since habitability could be taken as proportional to carrying capacity, as defined by the HSI, it is then related to the fraction of mass (*e.g.*, nutrients) and energy (*e.g.*, light) available or usable by a particular species or community from the environment. A common and difficult task of the HSIs is how to combine the effect of many environmental variables into a single index. The solutions are called aggregation methods in theoretical ecology. For example, these methods can combine the variables using arithmetic, geometric, or harmonic means, among others. The general rule is to keep the index proportional to carrying capacity and correlated with the presence and absence of the species of interest in the environment. Occurrences or presence probabilities are generally simpler to combine as products. Ecophysiological response curves often involve the fitting of standard statistical models to ecological data using simple (multiple) regression, Generalised Linear Models (GLM), Generalised Least Squares (GLS), or Generalised Additive Models (GAM), among others.

The usual approach is to create a library of HSI models for all species (or communities) and environments under consideration, each with its own particular limitations (Brooks, 1997; Roloff & Kernohan, 1999). These models are easy to compare and combine since they use the same uniform scale (*e.g.*, a value between zero and one, proportional to the carrying capacity). Thus, each HSI is only applicable to a specific type of life and habitat as a function of a finite set of environmental variables within selected spatial and temporal scales. There are many other tools of the HSM that can be used to characterize species or their environment. For example, *similarity indices* are usually simpler to construct than an HSI and can be used for quick comparisons between a set of biological or physical properties (*e.g.*, diversity) (Boyle *et al.*, 1990). Similarity indices are also used in many other applications such as pattern recognition and machine learning (*e.g.*, Cheng *et al.*, 2011).

## 3.  Habitability in Astrobiology: Proxies for Habitability

Astrobiologists have proposed many habitability models or indices for Earth, the Solar System, and extrasolar bodies in the last decade (*e.g.*, Stoker *et al.*, 2010; Schulze-Makuch *et al.*, 2011; Armstrong *et al.*, 2014; Barnes *et al.*, 2015; Silva *et al.*, 2017; Kashyap Jagadeesh *et al.*, 2017; Rodríguez-López *et al.*, 2019, Seales & Lenardic, 2020). There are some specific universal biological quantities that can be used as proxies for habitability such as carrying capacity, growth rate, metabolic rate, productivity, or the presence of some requirements of life, or even genetic diversity (Heller 2020). There is also an ongoing debate as to whether any concept of habitability needs to be binary (yes/no) in nature, as proposed by Cockell *et al.* (2019), or continuous, as opposed by Heller (2020), or probabilistic (Catling *et al.*, 2018). While a binary interpretation of habitability only allows a given planet to be habitable (to a given species) or not, a continuous model also allows for the possibility of a world (planet or moon) to be even more habitable than Earth, that is, to be superhabitable (Heller & Armstrong, 2014). Constructing a direct measure of habitability requires knowing how the environment affects one of the biological quantities for some species or community. We do not need to specifically estimate these quantities, but only to know how the environment proportionally affects them. For example, we know how temperature affects the productivity of primary producers such as plants and phytoplankton. Most require temperatures between 0° and 50° C, but such producers do better (*i.e.*, have the highest productivity) near 25° C (Silva et al., 2017). Their 'thermal habitability function' looks like a bell-shaped curve centered at their optimum productivity temperature. Direct measures of habitability are also better represented as a fraction from zero to one.

Biological productivity is the dry or carbon biomass produced over space and time. It is one of the best habitability proxies since it is easy to estimate for many ecosystems, via ground or satellite observations. The *Miami Model* was the first global-scale empirical model to give fair estimates of terrestrial net primary productivity (NPP, the rate of photosynthetic carbon fixed minus the carbon used by autotrophic respiration) (Zaks *et al*., 2007). This simple model only uses two measurements, annual mean surface temperature and precipitation, to successfully infer on the global distribution of vegetation (Adams *et al*., 2004). One important limitation of this type of models is that climate variables such as precipitation not only affect but are also affected by vegetation, *e.g.* there is increasing evidence that tropical forests have strong impacts on precipitation patterns on Earth (Molina *et al*., 2019). Today, many complex biogeochemical models and satellite observations (e.g., NASA's TERRA, AQUA, and Soumi NPP models) are combined to estimate local to global NPPs (Cramer *et al*., 1999; Ito, 2011). These satellite products are being used to create habitability indices to monitor terrestrial biodiversity now and through climate change (*e.g.*, Pan *et al*., 2010; Radeloff *et al*., 2019). Therefore, the NPP is also a measure of global terrestrial health or habitability since primary producers are the basis of the food chain.

Most habitability models are limited to indirect measures of habitability due to a lack of information. This is especially true for extrasolar planets (exoplanets). For example, the occurrence of Earth-size planets in the Habitable Zone of stars (termed the *Eta-Earth* value) can be considered a continuous indirect measure of stellar habitability (*i.e.*, the suitability of stars for habitable planets). The Habitable Zone, the region around a star where an Earth-like planet could maintain surface liquid water, is a general considered to be a binary indirect measure of planetary habitability (Kasting *et al*., 1993) (although others have argued that it should be considered a probability density function (Zsom, 2015; Catling *et al*., 2018). Although the location of the Habitable Zone depends mainly on the stellar type, its extension depends on the physical properties of the planet, in particular on the planet's atmospheric response to the stellar flow it receives (Kane, 2013). Thus, the presence of liquid water on the surface also depends on the planet's atmospheric dynamics, which effectively work to homogenize differential heating of the surface, creating a short-term response on the planet's global temperature. This differential heating is a result of the planet's obliquity, which governs the latitudinal distribution of incoming stellar radiation (Nowajewski *et al*., 2018).

The Habitable Zone can be defined in terms of either the planet's distance from the star, its incoming stellar flux, or its global equilibrium temperature. When using the equilibrium temperature definition, the extension of the Habitable Zone depends on the planet's orbital forcings, particularly eccentricity and obliquity. For example, when orbital eccentricity increases, the average equilibrium temperature decreases, thus extending the size of the Habitable Zone (Méndez & Rivera-Valentín, 2017). Similarly, higher fixed obliquity and/or rapid changes in obliquity values result in higher average equilibrium temperatures, which also result in extending the outer edge of the Habitable Zone (Armstrong *et al*., 2014). Further, when using the equilibrium temperature definition, the extension of the Habitable Zone depends ultimately on the planet's energy balance. On earth, the global energy balance is a result of the complex interaction between physical and biological processes. Biota affects the global energy balance in manifold ways including direct effects on surface albedo and latent heat fluxes (*e.g.*, transpiration) (Jasechko *et al.*, 2013).

The Earth Similarity Index (ESI), inspired by the diversity similarity indices used in ecology to compare populations (Boyle *et al.*, 1990), is a measure of Earth-likeness for a selected set of planetary parameters (Schulze-Makuch *et al.*, 2011). Future observational constraints of Earth-similar atmospheric constituents (*i.e.*, $N_2$, $CO_2$, $H_2O$) could improve our handle on this and similar metrics. For instance, 3D global climate models indicate that spectral features of water vapor on close-in terrestrial exoplanetary atmospheres may be detectable by the *James Webb Space Telescope* (Kopporapu *et al.*, 2017; Chen *et al.*, 2019), depending on the presence of clouds (Komacek *et al.*, 2020). Even though the presence of water vapor in the atmospheres of terrestrial exoplanets can indicate habitability, it is necessary to perform exhaustive work to determine which species could survive under conditions of extreme humidity. For example, mammals are not capable of surviving hyperthermia produced under high air temperatures and high humidity conditions, so planets with extreme differential heating between latitudes may be uninhabitable for them despite having liquid water on their surface (Nowajewski *et al.*, 2018).

The current Habitable Zone paradigm is misunderstood by many people — the public, the press, as well as other scientists — but, as all habitability models, it has a specific application and it is not incorrect or useless for neglecting the subsurface oceans in the outer Solar System, the Venus clouds, or other environments far from Earth-like conditions. The Habitable Zone does not tell us if the planets there are habitable (not even if there are planets there) but it shows the impact of a few important variables in planetary habitability. The concept of a Habitable Zone was developed to identify terrestrial exoplanet targets that could potentially host life. It was first proposed by Edward Maunder in 1913 (Maunder, 1913, Lorenz, 2020) in his book *Life on Other Planets* with refining definitions later on (Huang, 1959; Hart, 1978; Kasting *et al.*, 1993; Underwood *et al.*, 2003; Selsis *et al.*, 2007; Kaltenegger & Sasselov, 2011; Kopparapu *et al.*, 2013, 2014; Ramirez & Kaltenegger 2017, 2018). The general definition of the Habitable Zone that is currently being used is *the circumstellar region around a star where a terrestrial planet with a suitable atmosphere could host liquid water on its surface*. The insistence on the presence of liquid water on the surface is based on the fact that life on Earth requires liquid water to sustain. This definition is suitable only for remote observation of planets and does not consider any life which might exist at the subsurface. There is a reason for that: The search for life on exoplanets will rely on remote observations of atmospheres for the foreseeable future, lacking the luxury of in-situ measurements used in solar system planetary science. Therefore, identifying water in the atmosphere of planets (in addition to other biosignature relevant gases) is the only way to narrow down potential life-hosting targets, as subsurface life deep in the interior may not be able to modify the atmospheres of planets enough to be remotely detectable.

Abundance of liquid water in a planetary environment may be inherently unstable (Gorshkov et al., 2004), which leads to questions about the role of life in the definition of habitability itself (Zuluaga *et al.*, 2014). Thermodynamic disequilibrium may be one the most conspicuous signatures of a habitable (and inhabited) planet (Kleidon, 2012). One common problem with some (if not all) biological models is that they assume that climate (more generally, the full set of physical characteristics of the environment) is *a boundary condition* for life, *i.e.* that biological systems depend on climate but not the other way around. This premise is challenged by the fact that the observed state of the Earth system is the result of a complex and dynamic interaction between biological (*e.g.*, ecosystems) and physical (*e.g.*, climate) systems (Budyko, 1974; Gorshkov *et al*., 2000; Kleidon, 2012; Zuluaga *et al*., 2014). A critical question is how such state (which in thermodynamically unstable; consider *e.g.* the composition of Earth's atmosphere (Lenton, 1998; Kleidon, 2012)) can be maintained during eons (the span of Earth's life) despite variable (*e.g.*, solar luminosity) and sudden large external forcings (*e.g.*, asteroid impacts). The answer depends on the interactions between biological and physical systems on Earth. A planet might be habitable (its state becomes compatible with the presence of liquid water) during a given period of time just *by chance* (a random change in the planetary energy balance due to any combination of reasons), but long-term persistence of a habitable state indicates the existence of natural regulation mechanisms (Walker *et al*., 1981, Lenton, 1998; Gorshkov *et al*., 2000; Kleidon & Lorenz, 2004, Salazar & Poveda, 2009), *e.g.* how Earth has maintained its habitable state during around 4 billion years.

## 4. A General Mass-Energy Model for Habitability

The classical elements of air (gas), water (liquid), earth (solid), and fire (energy) are a powerful analogy to understand habitability. In essence, life needs environments where the three phases of matter and energy coexist. The five elements that constitute RNA and DNA (H, C, N, O, and P) are not all available in a single phase of matter. For example, under temperate conditions, nitrogen is more likely available in a gas phase, whereas inorganic sulfur is widely available in rocks as sulfates. The dissolution, diffusion, and flow capabilities of air and water allow all these elements to mix and be readily available in any habitat. Following this analogy, when a product of life, such as wood, is burned it releases all the classical elements back to the environment: gas (*e.g.*, $CO_2$), liquid (water), solid (ashes), and energy (heat). On a planetary scale, the classical elements become the atmosphere, hydrosphere, lithosphere — the habitable trinity of Dohm & Maruyama (2015) — plus photosynthetic or chemosynthetic energy. Thus, habitability depends fundamentally on the availability of raw materials and energy to assemble and maintain life, among other factors (e.g., decay and recycling of matter).

The analogy between habitability and the classical elements can be formalized into a general quantitative framework. Here we propose that *habitability is proportional to the mass and energy available for life*. This definition is consistent with the habitability models developed in ecology (*i.e.*, proportional to the carrying capacity) and explicitly includes energy as suggested by others (Hoehler, 2007; Macalady *et al*., 2013). Our model is applicable to any type of environment, from microenvironments to entire biospheres. The main challenge is to convert the intensive or extensive properties of interest (*e.g.*, temperature and water activity) to quantities proportional to the mass and energy of the environment. The mass available for life is usually a small fraction of the total environment mass and is further constrained by any limiting ingredient (*e.g.*, nutrients). For example, having more water does not make oceans more habitable, as life is limited by the availability of iron and sulfur in them. In fact, high concentrations of some ingredients might be harmful because they dilute other essential ingredients or make the environment toxic. Negative factors, like ionizing radiation, reduce the fraction of the available mass and energy (*e.g.*, Atri, 2017).

Our proposed habitability model can be constructed in six steps: (1) select the space and time of the region of interest (*i.e.*, define the boundary conditions); (2) select variables and convert them to quantities proportional to mass or energy; (3) select species or communities and their ecophysiological response curves for the selected variables; (4) identify one or more standards of comparison (*i.e.*, a terrestrial or planetary analog); and (5) solve the habitability master equation (Méndez *et al.*, in preparation):

$$\frac{\partial^2 H}{\partial s \partial t} = \left( \frac{1}{M} \frac{\partial^2 M}{\partial s \partial t} + \frac{1}{E} \frac{\partial^2 E}{\partial s \partial t} \right) H + \frac{\partial M}{\partial s} \frac{\partial E}{\partial t} + \frac{\partial M}{\partial t} \frac{\partial E}{\partial s} \qquad (1)$$

where *H* is the habitability, *M* and *E* are the normalized mass and energy available (*i.e.*, a fraction of the total mass and energy) for the life of interest (*i.e.*, a species or community) relative to the standard of comparison, respectively, and *s* and *t* are the spatial (*e.g.*, area or volume) and temporal (*e.g.*, hours or days) components. The last step, then, is to (6) validate and correct the habitability model with environments on Earth, if possible (*i.e.*, find positive correlations between habitability and biomass, productivity, or biosignatures). For example, equation 1 can be used to compare the habitability of a specific volume of ocean water of Europa relative to the same volume of deep ocean waters on Earth, given mass and energy fluxes. Sometimes it is not desirable to use the same volume or time periods for comparison purposes (*e.g.*, comparing early Mars with contemporary Earth). The space of interest is not limited by planetary scales. It can be enclosing a stellar region to evaluate its overall habitability, *e.g.*, a galactic habitable zone (Spitoni *et al.*, 2017). The general population growth equations (*e.g.*, exponential and logistic) can also be derived from equation 1.

The construction of a habitability model is not easy. Our model provides at least an upper limit for the habitability of a system for a given set of parameters, and is further improved by properly selecting key environmental variables and ecophysiological response curves for the life of interest (steps 2 and 3). Any model must be validated with environments on Earth where a positive correlation between habitability and the presence or abundance of life should be observed (step 6). The simple solution of equation 1 is *H* = *ME* (units of kgJ), or more practically, *H'* = $\rho P$ (units of Wkgm$^{-3}$), where *H'* is the specific habitability, $\rho$ is the concentration of one or more ingredients necessary for life, and *P* is the available metabolic power (assuming that mass only depends on space and energy on time). In practice, each of these variables could be normalized to the standard of comparison for simplicity and consistency, where zero denotes a non-habitable environment and one denotes a highly habitable environment. Also, occurrence or probabilities could be used instead of these variables, which is exactly what the definition of the Habitable Zone does (*i.e.*, probability of surface water ≥ 0). Negative values could be used to quantify the damaging effect of non-habitable environments (*e.g.*, comparing the surface of the Moon and Venus). Values larger than one could represent super-habitable conditions relative to a standard.

As an example, here are the six steps for the construction of a simple habitability model of the Martian surface at the landing site of the Mars Science Laboratory (MSL) rover in Gale Crater. Step 1: Our spatial and temporal regions of interest are the near-surface atmosphere diurnal cycles (*e.g.*, a cubic meter enclosing the atmosphere). Step 2: For simplicity, our habitability calculation will be constructed as a function of temperature and relative humidity with respect to liquid because these two variables are known to control microbial growth and have been measured in-situ. Thus, the temperature here is our proxy for the available energy while relative humidity is our proxy for the available mass, in this case water. Step 3: Here we are interested in microbial growth under Martian conditions, probably by some photosynthetic endolithic organisms similar to those in the Atacama Desert (Wierzchos *et al.*, 2013). There are many experiments on microbial growth as a function of both temperature and relative humidity (*e.g.*, McEldowney & Fletcher, 1988), which could be used to select suitable ecophysiology response curves. However, here we avoid this step to simplify our model. Step 4: Our analog for comparison is the Atacama Desert since we are interested in microbial growth under similar arid conditions. Although we note that the Atacama has orders of magnitude more water vapor than Mars and so is not directly comparable to the Martian surface. Step 5: Using the solution of equation 1, *H = ME*, the thermal-humidity habitability *H(T*,RH) is

$$H(T, \text{RH}) = \frac{ME}{[ME]_{\text{ref}}} = \frac{(q_m \bar{\rho} V)(q_e A t \sigma T^4)}{[(q_m \bar{\rho} V)(q_e A t \sigma T^4)]_{\text{ref}}} = \frac{\text{RH} \rho_e T^4}{[\text{RH} \rho_e T^4]_{\text{ref}}} \qquad (2)$$

where the denominator corresponds to the quantities of the comparison analog, the Atacama Desert. Here $q_m$ and $q_e$ are the quality factors, the usable fraction from the available mass and energy for life, respectively. Although these factors are unknown (unless controlled lab experiments are done), we can assume for simplicity that they are the same as the standard of comparison and cancel them out. The saturation vapor pressure for liquid water is given by $\rho_e$. Since we are comparing the same spatial and temporal scales, the volume *V*, area *A*, and time *t* also cancel out. Step 6: Data from the literature or new experiments could be used to validate equation 2. There should be a positive correlation between standing biomass and habitability. For example, we validated a similar expression with terrestrial biomes (Mendez *et al.*, 2018).

The thermal-humidity habitability of equation 2 was used to evaluate the habitability of the warmest and most humid sol, respectively (Figure 1). We can conclude from this simple analysis that the martian surface is about two to three orders of magnitude less habitable, in terms of temperature and relative humidity, than the Atacama Desert. This implies orders of magnitude less biologically produced organic carbon than the Atacama Desert if life were present. Hot sols are about twice more habitable than *wet* sols. Habitability peaks later in the afternoon of wet sols, around 16 LST, and two hours earlier in hot sols, around 14 LST. Habitable conditions are slightly longer in hot sols than in wet sols. This habitability analysis provides an upper limit to the habitability of mars for these particular sols, since we assumed (*i.e.*, due to the quality factors) that life there tolerates the environment and it is as efficient at extracting water and energy from the environment as microbial life in the Atacama Desert. Even so, without considering other factors (*e.g.*, UV), these types of models show how hostile the surface environment of Mars is compared to Earth (a *polite* way to say non-habitable).

Indeed, the likely time periods when liquids would be stable and forming on Mars do not coincide with the derived peak habitability values (Rivera-Valentin *et al.*, 2020). It is recommended that habitability models provide some value, even small but different from zero, to even compare hostile environments. There is no knowledge gained if a habitability model determines exactly zero habitability for all non-habitable environments (*e.g.*, Which environment is less habitable, the surface of Venus or Titan?). Besides, life *as we do not know it* could be thriving in conditions far from those expected for terrestrial life (National Research Council, 2007).

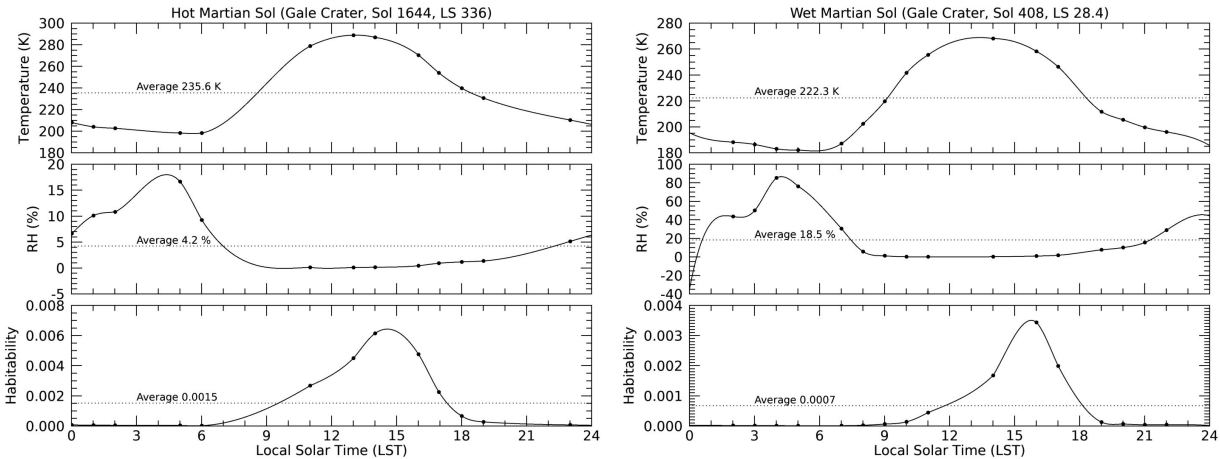

**Figure 1.** Near-surface temperature, relative humidity with respect to liquid, and thermal-humidity habitability for a hot (left) and wet (right) martian sol as measured by the Mars Science Laboratory Rover Environmental Monitoring Station. The thermal-humidity habitability was calculated from equation 2 relative to the Atacama Desert as the comparison standard. Temperature and relative humidity with respect to liquid data from Rivera-Valentín *et al.* (2018).

As in the previous example, our general mass-energy habitability could be used to calculate and compare the habitability of ocean worlds (*e.g.*, Europa and Enceladus), the lakes of Titan, or the clouds of Venus. This is not an Earth-centric model since mass and energy are conserved quantities and no life could take more mass and energy than available in the environment. The general modeling approach is to start first with simple models (e.g. fewer variables under steady-conditions) and then move forward with complex dynamical models. A library is then created depending on the variables, the scales (e.g, site, regional, global), or life forms of interest. Likewise, there are several interesting terrestrial analogs that would be interesting to look into with our habitability models. Our models might be particularly well suited to look at low diversity and low biomass extreme environments. Also interesting are extreme environments said to be sterile (*e.g.*, the acidic brines of Dallol-Danakil or brines with a very high content of Mg or other chaotropes) (Belilla *et al.*, 2019).

# 5. Recommendations for Planetary Exploration Missions

Planetary exploration missions are playing a critical role in our understanding of planetary habitability beyond what remote sensing from space can provide. The habitability of particular environments on planets such as Mars or Europa can be explored and compared thanks to targeted measurements taken with multiple orbital and ground sensors (*e.g.*, temperature, radiation, etc.), with which habitability models can be constructed. At the same time, future mission designs can synergistically take advantage of the predictions of habitability models in their selection of potential exploration strategies, mission priorities, and instruments, whether they are primarily astrobiological missions or not. Planetary exploration mission designs for the upcoming decades may have major astrobiological components that can directly or indirectly inform the study of habitability in the Solar System — even if the determination of habitability is not the primary focus of a mission. Indeed, general mission components not directly designed for astrobiological purposes might usefully contribute to habitability studies with only minimal considerations in design. Here we list three recommendations for the planetary community:

1. **Increase and widen the participation of more experts on habitat suitability models.** Ecologists are the experts in the ground-truthed proven measurement of terrestrial habitability, yet they are seldom represented in the planetary and astrobiology community. New synergies between NASA and the national and international ecological societies, *e.g.*, the Ecological Society of America (ESA), Soil Ecology Society (SES), and the International Society for Microbial Ecology (ISME), should be established via, for example, a joint conference session at the Lunar and Planetary Science Conference. There should be worldwide participation to guarantee global standardization. This synergy will stimulate the participation and exploration of the Solar System as a laboratory for expanding our current understanding of the habitability of Earth.

2. **Further terrestrial exploration.** Many Earth habitats are vastly under-explored biologically. For example, the clouds, stratosphere, deep ocean, deep ice, deep earth, or the mantle (*e.g.*, Lollar et al., 2019; DasSarma *et al*., 2020). Further, astrobiology needs to make stronger connections to the researchers working in these under-studied environments (*e.g.*, [The Deep Carbon Observatory](#)) so that there is a cohesive understanding of the state-of-the-art science being learned and efforts to continue to study these environments are supported. These field studies should provide new data to test the applicability of current habitability models with extreme environments, and thus get us closer to diverse planetary conditions. At the same time, unicellular life continually surprises us with new ways to survive and obtain energy from its environment (rock-eaters, electric currents, and even radioactivity) which shows us we need to be flexible in considering energy sources for habitability.

3. **Improve habitability models.** New habitability models should be developed and validated with field and laboratory experiments, including simulated extreme and planetary analog environments (*e.g.*, Taubner *et al*., 2020). The main goal is to identify knowledge gaps. For example, new ecophysiological response curves (*e.g.*, growth rate as a function of water activity, a measure of available water) for some organisms are necessary, especially in dynamic environments such as gradient-rich biotopes and higher complexity extreme environments (*i.e.*, those with multiple extremes such as deep-sea brines). Also, there are insufficient models on microbial growth in near-surface dynamic environments (*e.g.*, as applicable to martian diurnal cycles). There is a growing body of literature about the manifold mechanisms through which life affects the Earth's climate system, including the global energy balance and atmospheric composition and dynamics. Advances in the understanding of climate-life interactions in the Earth System (e.g. Bonan and Doney, 2018) can provide new insights for habitability models.

4. **Develop a NASA Habitability Standard (NHS).** Existing and future planetary missions should specify how they assess habitability for each of their instruments according to a shared NASA habitability standard. For example, measurements of surface temperature and water vapor from landers or orbital missions could be converted into a simple habitability model. The advantage of a standard is that past and future missions could be compared to each other and their habitability assessments refined, and new habitability knowledge gaps could be identified. This dynamic standard should be evaluated and updated regularly by a diverse and multidisciplinary committee, for example during a Decadal Survey and/or mid-decade review. Currently, the closest concept to an NHS is specific language included in various NASA roadmaps, such as the NASA Astrobiology Roadmap (Des Marais *et al.*, 2008) and the NASA Roadmap to Ocean Worlds (Hendrix *et al*., 2018). These documents stress the need for habitability evaluations and missions (*e.g.,* Europa Clipper and Titan Dragonfly), yet only focus on the individual habitability requirements and not how to combine the net contribution of these factors. Furthermore, the NHS might eventually become the standard of other disciplines.

## 6. Science Questions

Each astrobiological relevant planetary mission should answer a series of basic scientific questions about the environment(s) to be studied as a core part of the planning process. The answers to these questions should be updated based on mission results. To do so, it is important to define an environment of interest, both in space and time (termed a *quadrat* in ecology), and anticipate the following science questions as part of the initial analysis:

1. **What are the limiting factors?** Usually, there is a small set of main factors (*e.g.*, edaphic factors) that influence living organisms (*e.g.*, water, nutrients). These will be the first set of variables to be used for the construction of a habitability model, which will later be refined with more variables. For example, primary productivity is mainly driven by temperature, precipitation, and nutrients on land, and by temperature and nutrient concentrations in the oceans, among other factors. In general, these factors should be directly or indirectly related to the mass and energy of the environment (*e.g.*, Martiny *et al.*, 2006; Pikuta *et al*., 2007; Williams & Hallsworth, 2009; Harrison *et al*., 2013; McKay, 2014; Lynch & Neufeld, 2015; Tecon & Or, 2017).

2. **What are the terrestrial and planetary analogs?** Identify at least one analog on Earth and one close planetary analog as the comparison standards (*i.e.*, for model normalizations). For example, if studying a particular martian environment, select the terrestrial polar deserts and a martian analog based on the variables of interest. The cross-comparison of similar types of environment (*e.g.*, salterns), as well as slightly different settings (*e.g.*, high salinity biotopes with different pH, temperature, or chemical conditions), could also prove useful. The subsurface oceans of Europa or Enceladus could be compared with deep seawater, hydrothermal systems, or deep-sea brines (Antunes *et al.*, 2020). Planetary atmospheres could be compared with high altitude or near-space regions. An analysis of similarities (*e.g.*, ANOSIM) could be used to formally select and compare these regions (Clarke, 1993).

3. **What is the habitability value?** The habitability of the region of interest is evaluated based on the selected environmental factors, and then compared with the selected Earth and planetary analogs, using a normalized scale from zero to one for simplicity. A library of habitability measures is usually constructed (*i.e.*, a habitability matrix), each for different considerations (*e.g.*, species). These inputs are then used to construct multivariate habitability maps (*niche quantification* in ecology) for site selections.

4. **What is the potential biomass?** The upper limits of biomass can be predicted based on the fluxes of mass and energy available for life, and usually a very small fraction of the total mass and energy. For example, biomass could be estimated from the available metabolic energy using the Metabolic Theory of Ecology (van der Meer, 2006; Schramski *et al.*, 2015; Clarke, 2017). These upper limits are used in the sensitivity designs of life detection experiments. Available free energy from known disequilibria has been used to estimate an upper limit on the biomass in the subsurface of Mars and its value depends on uncertainties of the abundances of metabolic reactants and the assumed microbial basal power requirement (Sholes *et al.*, 2019).

5. **What is the expected correlation between habitability and biosignatures?** The potential upper values of biomass can be converted to estimates of observable biosignatures or disequilibrium chemistry (Catling *et al.*, 2018). Habitability and biosignatures are positively correlated on Earth but this might not be necessarily true for other planets. A zero or negative correlation could indicate an incorrect habitability model or a biological process unlike Earth (*life as we don't know it*). The habitability-biosignatures correlation is a fundamental problem of astrobiology, but non-detections are also important. For example, it will be profound to detect planetary regions determined to be habitable by Earth standards yet devoid of any life. Such discoveries would place bounds on abiogenesis.

## 7. Conclusion

Habitability models are successful analysis tools for characterizing habitable environments on Earth. Ecologists have been using these models for more than four decades to understand the distribution of terrestrial life at local to global scales (Section 2). Astrobiologists have been proposing different models for some time, with little integration and consistency between them and different in function to those used by biologists (Section 3). In this white paper, we suggest a mass-energy habitability model as an example of how to adapt and expand the models used by ecologists to the astrobiology field (Section 4). Our model could be used to compare environments and prioritize targets for exploration. NASA should create habitability standards for planetary missions with astrobiology objectives, as the USFWS successfully did long ago for ecologists (Section 5). These standards are necessary to make sense of data from multiple missions, develop predictions for environmental niches on planetary bodies that can be tested, and understand the extraterrestrial correlations between habitability and biosignatures (Section 6).

There is no need for the planetary and astrobiology community to reinvent the methods and tools used by ecologists. It is true now that the ecology methods are more capable than our limited planetary data allows, but they provide the basic language and framework to connect Earth and planetary sciences for decades to come. For example, there are many theoretical and computational tools used in ecology to quantify environments and their habitability, mostly known as habitat suitability models. See Guisan *et al*. (2017) for an extensive review of these models and Lortie *et al*. (2020) for a current review of the computational tools. Most of these tools are available as packages in the [R Computing Language](#) in the [Comprehensive R Archive Network (CRAN)](#) and [GitHub](#) (*e.g.*, [Environmetrics](#), [HSDM](#)). New, higher-resolution remote sensing instruments and exploration technologies will create better habitability maps from rover, lander, and orbiter data. Habitability models will eventually lead us to a better understanding of the potential for life in the Solar System and beyond, and perhaps even the factors that influence the development of life itself. *Habitability models are the foundation of planetary habitability science*. After all of our scientific and technological advances, we still need a stronger integration between biology, planetary sciences, and astronomy (Cockell, 2020).

**Acknowledgments**

This work was supported by the NASA Astrobiology Institute (NAI), the Planetary Habitability Laboratory (PHL), and the University of Puerto Rico at Arecibo (UPR Arecibo). Thanks to NASA Puerto Rico Space Grant Consortium and the Puerto Rico Louis Stokes Alliance For Minority Participation (PR-LSAMP) for supporting our students. Thanks to Ravi kumar Kopparapu from the NASA Goddard Space Flight Center for valuable comments.

## 8. References

Adams, B., White, A., & Lenton, T. M. (2004). An analysis of some diverse approaches to modelling terrestrial net primary productivity. *Ecological Modelling*, *177*(3), 353–391. https://doi.org/10.1016/j.ecolmodel.2004.03.014

Antunes, A., Olsson-Francis, K., & McGenity, T. J. (2020). Exploring Deep-Sea Brines as Potential Terrestrial Analogues of Oceans in the Icy Moons of the Outer Solar System. *Current Issues in Molecular Biology*, 123–162. https://doi.org/10.21775/cimb.038.123

Armstrong, J. c., Barnes, R., Domagal-Goldman, S., Breiner, J., Quinn, T. r., & Meadows, V. s. (2014). Effects of Extreme Obliquity Variations on the Habitability of Exoplanets. *Astrobiology*, *14*(4), 277–291. https://doi.org/10.1089/ast.2013.1129

Atri, D. (2017). Modelling stellar proton event-induced particle radiation dose on close-in exoplanets. *Monthly Notices of the Royal Astronomical Society: Letters*, *465*(1), L34–L38. https://doi.org/10.1093/mnrasl/slw199

Austin, M. P., & Gaywood, M. J. (1994). Current problems of environmental gradients and species response curves in relation to continuum theory. *Journal of Vegetation Science*, *5*(4), 473–482. https://doi.org/10.2307/3235973

Barnes, R., Meadows, V. S., & Evans, N. (2015). COMPARATIVE HABITABILITY OF TRANSITING EXOPLANETS. *The Astrophysical Journal*, *814*(2), 91. https://doi.org/10.1088/0004-637X/814/2/91

Belilla, J., Moreira, D., Jardillier, L., Reboul, G., Benzerara, K., López-García, J. M., Bertolino, P., López-Archilla, A. I., & López-García, P. (2019). Hyperdiverse archaea near life limits at the polyextreme geothermal Dallol area. *Nature Ecology & Evolution*, *3*(11), 1552–1561. https://doi.org/10.1038/s41559-019-1005-0

Bonan, G. B., & Doney, S. C. (2018). Climate, ecosystems, and planetary futures: The challenge to predict life in Earth system models. *Science*, *359*(6375). https://doi.org/10.1126/science.aam8328

Boyle, T. P., Smillie, G. M., Anderson, J. C., & Beeson, D. R. (1990). A Sensitivity Analysis of Nine Diversity and Seven Similarity Indices. *Research Journal of the Water Pollution Control Federation*, *62*(6), 749–762. JSTOR.

Brooks, R. P. (1997). Improving Habitat Suitability Index Models. *Wildlife Society Bulletin (1973-2006)*, *25*(1), 163–167. JSTOR.

Budyko (Ed.). (1974). *Climate and life* (English ed edition). Academic Press.

Cárdenas, R., Nodarse-Zulueta, R., Perez, N., Avila-Alonso, D., & Martin, O. (2019). On the Quantification of Habitability: Current Approaches. In R. Cárdenas, V. Mochalov, O. Parra, & O. Martin (Eds.), *Proceedings of the 2nd International Conference on BioGeoSciences* (pp. 1–8). Springer International Publishing. https://doi.org/10.1007/978-3-030-04233-2_1

Cardenas, R., Perez, N., Martinez-Frias, J., & Martin, O. (2014). On the Habitability of Aquaplanets. *Challenges*, *5*(2), 284–293. https://doi.org/10.3390/challe5020284

Catling, D. C., Krissansen-Totton, J., Kiang, N. Y., Crisp, D., Robinson, T. D., DasSarma, S., Rushby, A. J., Del Genio, A., Bains, W., & Domagal-Goldman, S. (2018). Exoplanet Biosignatures: A Framework for Their Assessment. *Astrobiology*, *18*(6), 709–738. https://doi.org/10.1089/ast.2017.1737

Chen, H., Wolf, E. T., Zhan, Z., & Horton, D. E. (2019). Habitability and Spectroscopic Observability of Warm M-dwarf Exoplanets Evaluated with a 3D Chemistry-Climate Model. *The Astrophysical Journal*, *886*(1), 16. https://doi.org/10.3847/1538-4357/ab4f7e

Cheng, F., Shen, J., Yu, Y., Li, W., Liu, G., Lee, P. W., & Tang, Y. (2011). In silico prediction of Tetrahymena pyriformis toxicity for diverse industrial chemicals with substructure pattern recognition and machine learning methods. *Chemosphere*, *82*(11), 1636–1643. https://doi.org/10.1016/j.chemosphere.2010.11.043

Chopra, A., & Lineweaver, C. H. (2016). The Case for a Gaian Bottleneck: The Biology of Habitability. *Astrobiology*, *16*(1), 7–22. https://doi.org/10.1089/ast.2015.1387

Clarke, A. (2017). The Metabolic Theory of Ecology. In *Principles of Thermal Ecology: Temperature, Energy, and Life*. Oxford University Press. https://www.oxfordscholarship.com/view/10.1093/oso/9780199551668.001.0001/oso-9780199551668-chapter-12

Clarke, K. R. (1993). Non-parametric multivariate analyses of changes in community structure. *Australian Journal of Ecology*, *18*(1), 117–143. https://doi.org/10.1111/j.1442-9993.1993.tb00438.x

Cockell, C. S. (2020). Astronomy + biology. *Astronomy & Geophysics*, *61*(3), 3.28-3.32. https://doi.org/10.1093/astrogeo/ataa042

Cockell, C. s., Bush, T., Bryce, C., Direito, S., Fox-Powell, M., Harrison, J. p., Lammer, H., Landenmark, H., Martin-Torres, J., Nicholson, N., Noack, L., O'Malley-James, J., Payler, S. j., Rushby, A., Samuels, T., Schwendner, P., Wadsworth, J., & Zorzano, M. p. (2016). Habitability: A Review. *Astrobiology*, *16*(1), 89–117. https://doi.org/10.1089/ast.2015.1295

Cockell, C. S., Stevens, A. H., & Prescott, R. (2019). Habitability is a binary property. *Nature Astronomy*, *3*(11), 956–957. https://doi.org/10.1038/s41550-019-0916-7

Cramer, W., Kicklighter, D. W., Bondeau, A., Iii, B. M., Churkina, G., Nemry, B., Ruimy, A., Schloss, A. L., & Intercomparison, T. P. O. T. P. N. M. (1999). Comparing global models of terrestrial net primary productivity (NPP): Overview and key results. *Global Change Biology*, *5*(S1), 1–15. https://doi.org/10.1046/j.1365-2486.1999.00009.x

DasSarma, P., Antunes, A., Simões, M. F., & DasSarma, S. (2020). Earth's Stratosphere and Microbial Life. *Current Issues in Molecular Biology*, 197–244. https://doi.org/10.21775/cimb.038.197

Des Marais, D. J., Nuth, J. A., Allamandola, L. J., Boss, A. P., Farmer, J. D., Hoehler, T. M., Jakosky, B. M., Meadows, V. S., Pohorille, A., Runnegar, B., & Spormann, A. M. (2008). The NASA Astrobiology Roadmap. *Astrobiology*, *8*(4), 715–730. https://doi.org/10.1089/ast.2008.0819

Dohm, J. M., & Maruyama, S. (2015). Habitable Trinity. *Geoscience Frontiers*, *6*(1), 95–101. https://doi.org/10.1016/j.gsf.2014.01.005

Douglas, A. E. (2018). What will it take to understand the ecology of symbiotic microorganisms? *Environmental Microbiology*, *20*(6), 1920–1924. https://doi.org/10.1111/1462-2920.14123

Farmer, J. D. (2018). Chapter 1—Habitability as a Tool in Astrobiological Exploration. In N. A. Cabrol & E. A. Grin (Eds.), *From Habitability to Life on Mars* (pp. 1–12). Elsevier. https://doi.org/10.1016/B978-0-12-809935-3.00002-5

Giles, R. H. (1978). *Wildlife Management*. W. H. Freeman.

Gorshkov, V. G., Makarieva, A. M., & Gorshkov, V. V. (2004). Revising the fundamentals of ecological knowledge: The biota–environment interaction. *Ecological Complexity*, *1*(1), 17–36. https://doi.org/10.1016/j.ecocom.2003.09.002

Gorshkov, V., Makarieva, A. M., & Gorshkov, V. V. (2000). *Biotic Regulation of the Environment: Key Issues of Global Change*. Springer Science & Business Media.

Guisan, A., Thuiller, W., & Zimmermann, N. E. (2017). *Habitat Suitability and Distribution Models: With Applications in R*. Cambridge University Press.

Harrison, J. P., Gheeraert, N., Tsigelnitskiy, D., & Cockell, C. S. (2013). The limits for life under multiple extremes. *Trends in Microbiology*, *21*(4), 204–212. https://doi.org/10.1016/j.tim.2013.01.006

Hart, M. H. (1978). The evolution of the atmosphere of the earth. *Icarus*, *33*, 23–39. https://doi.org/10.1016/0019-1035(78)90021-0

Heller, R. (2020). Habitability is a continuous property of nature. *Nature Astronomy*, *4*(4), 294–295. https://doi.org/10.1038/s41550-020-1063-x

Heller, R., & Armstrong, J. (2013). Superhabitable Worlds. *Astrobiology*, *14*(1), 50–66. https://doi.org/10.1089/ast.2013.1088

Hendrix, A. R., Hurford, T. A., Barge, L. M., Bland, M. T., Bowman, J. S., Brinckerhoff, W., Buratti, B. J., Cable, M. L., Castillo-Rogez, J., Collins, G. C., Diniega, S., German, C. R., Hayes, A. G., Hoehler, T., Hosseini, S., Howett, C. J. A., McEwen, A. S., Neish, C. D., Neveu, M., … Vance, S. D. (2018). The NASA Roadmap to Ocean Worlds. *Astrobiology*, *19*(1), 1–27. https://doi.org/10.1089/ast.2018.1955

Hirzel, A. H., & Lay, G. L. (2008). Habitat suitability modelling and niche theory. *Journal of Applied Ecology*, *45*(5), 1372–1381. https://doi.org/10.1111/j.1365-2664.2008.01524.x

Hoehler, T. M. (2007). An Energy Balance Concept for Habitability. *Astrobiology*, *7*(6), 824–838. https://doi.org/10.1089/ast.2006.0095

Huang, S.-S. (1959). The Problem of Life in the Universe and the Mode of Star Formation. *Publications of the Astronomical Society of the Pacific*, *71*, 421. https://doi.org/10.1086/127417

Ito, A. (2011). A historical meta-analysis of global terrestrial net primary productivity: Are estimates converging? *Global Change Biology*, *17*(10), 3161–3175. https://doi.org/10.1111/j.1365-2486.2011.02450.x

Jasechko, S., Sharp, Z. D., Gibson, J. J., Birks, S. J., Yi, Y., & Fawcett, P. J. (2013). Terrestrial water fluxes dominated by transpiration. *Nature*, *496*, 347–350. https://doi.org/10.1038/nature11983

Kaltenegger, L., & Sasselov, D. (2011). Exploring the Habitable Zone for Kepler Planetary Candidates. *The Astrophysical Journal Letters*, *736*, L25. https://doi.org/10.1088/2041-8205/736/2/L25

Kane, S. R. (2013). The Habitable Zone: Basic Concepts. In J.-P. de Vera & J. Seckbach (Eds.), *Habitability of Other Planets and Satellites* (pp. 3–12). Springer Netherlands. https://doi.org/10.1007/978-94-007-6546-7_1

Kashyap Jagadeesh, M., Gudennavar, S. B., Doshi, U., & Safonova, M. (2017). Indexing of exoplanets in search for potential habitability: Application to Mars-like worlds. *Astrophysics and Space Science*, *362*(8), 146. https://doi.org/10.1007/s10509-017-3131-y

Kasting, J. F., Whitmire, D. P., & Reynolds, R. T. (1993). Habitable Zones around Main Sequence Stars. *Icarus*, *101*(1), 108–128. https://doi.org/10.1006/icar.1993.1010

Kleidon, A. (2012). How does the Earth system generate and maintain thermodynamic disequilibrium and what does it imply for the future of the planet? *Philosophical Transactions of the Royal Society A: Mathematical, Physical and Engineering Sciences*, *370*(1962), 1012–1040. https://doi.org/10.1098/rsta.2011.0316

Kleidon, A., & Lorenz, R. D. (2004). *Non-equilibrium Thermodynamics and the Production of Entropy: Life, Earth, and Beyond*. Springer Science & Business Media.

Komacek, T. D., Fauchez, T. J., Wolf, E. T., & Abbot, D. S. (2020). Clouds will Likely Prevent the Detection of Water Vapor in JWST Transmission Spectra of Terrestrial Exoplanets. *The Astrophysical Journal*, *888*(2), L20. https://doi.org/10.3847/2041-8213/ab6200

Kopparapu, R. kumar, Wolf, E. T., Arney, G., Batalha, N. E., Haqq-Misra, J., Grimm, S. L., & Heng, K. (2017). Habitable Moist Atmospheres on Terrestrial Planets near the Inner Edge of the Habitable Zone around M Dwarfs. *The Astrophysical Journal*, *845*(1), 5. https://doi.org/10.3847/1538-4357/aa7cf9

Kopparapu, R. K., Ramirez, R., Kasting, J. F., Eymet, V., Robinson, T. D., Mahadevan, S., Terrien, R. C., Domagal-Goldman, S., Meadows, V., & Deshpande, R. (2013). Habitable Zones around Main-sequence Stars: New Estimates. *The Astrophysical Journal*, *765*, 131. https://doi.org/10.1088/0004-637X/765/2/131

Kopparapu, R. K., Ramirez, R. M., SchottelKotte, J., Kasting, J. F., Domagal-Goldman, S., & Eymet, V. (2014). Habitable Zones around Main-sequence Stars: Dependence on

Planetary Mass. *The Astrophysical Journal Letters*, *787*, L29. https://doi.org/10.1088/2041-8205/787/2/L29

Kuhn, T., Cunze, S., Kochmann, J., & Klimpel, S. (2016). Environmental variables and definitive host distribution: A habitat suitability modelling for endohelminth parasites in the marine realm. *Scientific Reports*, *6*(1), 30246. https://doi.org/10.1038/srep30246

Lenton, T. M. (1998). Gaia and natural selection. *Nature*, *394*(6692), 439–447. https://doi.org/10.1038/28792

Lollar, G. S., Warr, O., Telling, J., Osburn, M. R., & Lollar, B. S. (2019). 'Follow the Water': Hydrogeochemical Constraints on Microbial Investigations 2.4 km Below Surface at the Kidd Creek Deep Fluid and Deep Life Observatory. *Geomicrobiology Journal*, *36*(10), 859–872. https://doi.org/10.1080/01490451.2019.1641770

Lorenz, R. D. (2020). Maunder's Work on Planetary Habitability in 1913: Early Use of the term "Habitable Zone" and a "Drake Equation" Calculation. *Research Notes of the AAS*, *4*(6), 79. https://doi.org/10.3847/2515-5172/ab9831

Lortie, C. J., Braun, J., Filazzola, A., & Miguel, F. (2020). A checklist for choosing between R packages in ecology and evolution. *Ecology and Evolution*, *10*(3), 1098–1105. https://doi.org/10.1002/ece3.5970

Lynch, M. D. J., & Neufeld, J. D. (2015). Ecology and exploration of the rare biosphere. *Nature Reviews Microbiology*, *13*(4), 217–229. https://doi.org/10.1038/nrmicro3400

Macalady, J. L., Hamilton, T. L., Grettenberger, C. L., Jones, D. S., Tsao, L. E., & Burgos, W. D. (2013). Energy, ecology and the distribution of microbial life. *Philosophical Transactions of the Royal Society B: Biological Sciences*, *368*(1622), 20120383. https://doi.org/10.1098/rstb.2012.0383

Martinez-Frias, J., Lázaro, E., & Esteve-Núñez, A. (2007). Geomarkers versus Biomarkers: Paleoenvironmental and Astrobiological Significance. *AMBIO: A Journal of the Human Environment*, *36*(5), 425–426. https://doi.org/10.1579/0044-7447(2007)36[425:GVBPAA]2.0.CO;2

Martiny, J. B. H., Bohannan, B. J. M., Brown, J. H., Colwell, R. K., Fuhrman, J. A., Green, J. L., Horner-Devine, M. C., Kane, M., Krumins, J. A., Kuske, C. R., Morin, P. J., Naeem, S., Øvreås, L., Reysenbach, A.-L., Smith, V. H., & Staley, J. T. (2006). Microbial biogeography: Putting microorganisms on the map. *Nature Reviews Microbiology*, *4*(2), 102–112. https://doi.org/10.1038/nrmicro1341

Maunder, E. W. (1913). Are the planets inhabited? *London, New York, Harper & Brothers, 1913*. http://adsabs.harvard.edu/abs/1913api..book.....M

McEldowney, S., & Fletcher, M. (1988). The effect of temperature and relative humidity on the survival of bacteria attached to dry solid surfaces. *Letters in Applied Microbiology*, *7*(4), 83–86. https://doi.org/10.1111/j.1472-765X.1988.tb01258.x

McKay, C. P. (2014). Requirements and limits for life in the context of exoplanets. *Proceedings of the National Academy of Sciences*, *111*(35), 12628–12633. https://doi.org/10.1073/pnas.1304212111

Méndez, A., & Rivera-Valentín, E. G. (2017). The Equilibrium Temperature of Planets in Elliptical Orbits. *The Astrophysical Journal*, *837*(1), L1. https://doi.org/10.3847/2041-8213/aa5f13

Méndez, A., Schulze-Makuch, D., Nery, G., Rivera-Valentin, E. G., Davila, A., Ramirez, R., Wood, T., Rodriguez-Garcia, A. D., Soto-Soto, A., Gonzalez-Villanueva, S. E., Rivera-Saavedra, S. M., Maldonado-Vazquez, G. J., Colon-Acosta, E., Cruz-Mendoza, V. M., Crespo Sanchez, J. K., & Estevez-Mesa, N. (2018). *A General Mass-Energy Habitability Model*. *49*, 2511.

Molina, R. D., Salazar, J. F., Martínez, J. A., Villegas, J. C., & Arias, P. A. (2019). Forest-Induced Exponential Growth of Precipitation Along Climatological Wind Streamlines Over the Amazon. *Journal of Geophysical Research: Atmospheres*, *124*(5), 2589–2599. https://doi.org/10.1029/2018JD029534

National Research Council. (2007). *The Limits of Organic Life in Planetary Systems*. https://doi.org/10.17226/11919

Nowajewski, P., Rojas, M., Rojo, P., & Kimeswenger, S. (2018). Atmospheric dynamics and habitability range in Earth-like aquaplanets obliquity simulations. *Icarus*, *305*, 84–90. https://doi.org/10.1016/j.icarus.2018.01.002

Oren, A. (1999). Bioenergetic Aspects of Halophilism. *Microbiology and Molecular Biology Reviews*, *63*(2), 334–348. https://doi.org/10.1128/MMBR.63.2.334-348.1999

Oren, A. (2001). The bioenergetic basis for the decrease in metabolic diversity at increasing salt concentrations: Implications for the functioning of salt lake ecosystems. In J. M. Melack, R. Jellison, & D. B. Herbst (Eds.), *Saline Lakes: Publications from the 7th International Conference on Salt Lakes, held in Death Valley National Park, California, U.S.A., September 1999* (pp. 61–72). Springer Netherlands. https://doi.org/10.1007/978-94-017-2934-5_6

Pan, Y., Cheong, C. M., & Blevis, E. (2010). The climate change habitability index. *Interactions*, *17*(6), 29–33. https://doi.org/10.1145/1865245.1865253

Pikuta, E. V., Hoover, R. B., & Tang, J. (2007). Microbial Extremophiles at the Limits of Life. *Critical Reviews in Microbiology*, *33*(3), 183–209. https://doi.org/10.1080/10408410701451948

Radeloff, V. C., Dubinin, M., Coops, N. C., Allen, A. M., Brooks, T. M., Clayton, M. K., Costa, G. C., Graham, C. H., Helmers, D. P., Ives, A. R., Kolesov, D., Pidgeon, A. M., Rapacciuolo, G., Razenkova, E., Suttidate, N., Young, B. E., Zhu, L., & Hobi, M. L. (2019). The Dynamic Habitat Indices (DHIs) from MODIS and global biodiversity. *Remote Sensing of Environment*, *222*, 204–214. https://doi.org/10.1016/j.rse.2018.12.009

Rajakaruna, N., & Boyd, R. S. (2008). Edaphic Factor. In S. E. Jørgensen & B. D. Fath (Eds.), *Encyclopedia of Ecology* (pp. 1201–1207). Academic Press. https://doi.org/10.1016/B978-008045405-4.00484-5

Ramirez, R. M., & Kaltenegger, L. (2017). A Volcanic Hydrogen Habitable Zone. *The Astrophysical Journal Letters*, *837*, L4. https://doi.org/10.3847/2041-8213/aa60c8

Ramirez, R. M., & Kaltenegger, L. (2018). A Methane Extension to the Classical Habitable Zone. *The Astrophysical Journal*, *858*, 72. https://doi.org/10.3847/1538-4357/aab8fa

Rivera-Valentín, E. G., Chevrier, V. F., Soto, A., & Martínez, G. (2020). Distribution and habitability of (meta)stable brines on present-day Mars. *Nature Astronomy*. https://doi.org/10.1038/s41550-020-1080-9

Rivera-Valentín, E. G., Gough, R. V., Chevrier, V. F., Primm, K. M., Martínez, G. M., & Tolbert, M. (2018). Constraining the Potential Liquid Water Environment at Gale Crater, Mars. *Journal of Geophysical Research (Planets)*, *123*, 1156–1167. https://doi.org/10.1002/2018JE005558

Rodríguez-López, L., Cardenas, R., Parra, O., González-Rodríguez, L., Martin, O., & Urrutia, R. (2019). On the quantification of habitability: Merging the astrobiological and ecological schools. *International Journal of Astrobiology*, *18*(5), 412–415. https://doi.org/10.1017/S1473550418000344

Roloff, G. J., & Kernohan, B. J. (1999). Evaluating Reliability of Habitat Suitability Index Models. *Wildlife Society Bulletin (1973-2006)*, *27*(4), 973–985. JSTOR.

Salazar, J. F., & Poveda, G. (2009). Role of a simplified hydrological cycle and clouds in regulating the climate–biota system of Daisyworld. *Tellus B: Chemical and Physical Meteorology*, *61*(2), 483–497. https://doi.org/10.1111/j.1600-0889.2009.00411.x

Schramski, J. R., Dell, A. I., Grady, J. M., Sibly, R. M., & Brown, J. H. (2015). Metabolic theory predicts whole-ecosystem properties. *Proceedings of the National Academy of Sciences of the United States of America*, *112*(8), 2617–2622. https://doi.org/10.1073/pnas.1423502112

Schulze-Makuch, D., Méndez, A., Fairén, A. G., von Paris, P., Turse, C., Boyer, G., Davila, A. F., António, M. R. de S., Catling, D., & Irwin, L. N. (2011). A Two-Tiered Approach to Assessing the Habitability of Exoplanets. *Astrobiology*, *11*(10), 1041–1052. https://doi.org/10.1089/ast.2010.0592

Seales, J., & Lenardic, A. (2020). Uncertainty Quantification in Planetary Thermal History Models: Implications for Hypotheses Discrimination and Habitability Modeling. *The Astrophysical Journal*, *893*(2), 114. https://doi.org/10.3847/1538-4357/ab822b

Selsis, F., Kasting, J. F., Levrard, B., Paillet, J., Ribas, I., & Delfosse, X. (2007). Habitable planets around the star Gliese 581? *Astronomy and Astrophysics*, *476*, 1373–1387. https://doi.org/10.1051/0004-6361:20078091

Shock, E. L., & Holland, M. E. (2007). Quantitative Habitability. *Astrobiology*, *7*(6), 839–851. https://doi.org/10.1089/ast.2007.0137

Sholes, S. F., Krissansen-Totton, J., & Catling, D. C. (2019). A Maximum Subsurface Biomass on Mars from Untapped Free Energy: CO and H2 as Potential Antibiosignatures. *Astrobiology*, *19*, 655–668. https://doi.org/10.1089/ast.2018.1835

Silva, L., Vladilo, G., Schulte, P. M., Murante, G., & Provenzale, A. (2017). From climate models to planetary habitability: Temperature constraints for complex life. *International Journal of Astrobiology*, *16*(3), 244–265. https://doi.org/10.1017/S1473550416000215

Spitoni, E., Gioannini, L., & Matteucci, F. (2017). Galactic habitable zone around M and FGK stars with chemical evolution models that include dust. *Astronomy and Astrophysics*, *605*, A38. https://doi.org/10.1051/0004-6361/201730545

Stoker, C. R., Zent, A., Catling, D. C., Douglas, S., Marshall, J. R., Archer, D., Clark, B., Kounaves, S. P., Lemmon, M. T., Quinn, R., Renno, N., Smith, P. H., & Young, S. M. M. (2010). Habitability of the Phoenix landing site. *Journal of Geophysical Research: Planets*, *115*(E6). https://doi.org/10.1029/2009JE003421

Taubner, R.-S., Olsson-Francis, K., Vance, S. D., Ramkissoon, N. K., Postberg, F., de Vera, J.-P., Antunes, A., Camprubi Casas, E., Sekine, Y., Noack, L., Barge, L., Goodman, J., Jebbar, M., Journaux, B., Karatekin, Ö., Klenner, F., Rabbow, E., Rettberg, P., Rückriemen-Bez, T., … Soderlund, K. M. (2020). Experimental and Simulation Efforts in the Astrobiological Exploration of Exooceans. *Space Science Reviews*, *216*(1), 9. https://doi.org/10.1007/s11214-020-0635-5

Tecon, R., & Or, D. (2017). Biophysical processes supporting the diversity of microbial life in soil. *FEMS Microbiology Reviews*, *41*(5), 599–623. https://doi.org/10.1093/femsre/fux039

U. S. Fish and Wildlife Service Division of Ecological. (1980). *Ecological services manual*. Division of Ecological Services, U.S. Fish and Wildlife Service, Department of the Interior.

Underwood, D. R., Jones, B. W., & Sleep, P. N. (2003). The evolution of habitable zones during stellar lifetimes and its implications on the search for extraterrestrial life.

*International Journal of Astrobiology*, *2*, 289–299. https://doi.org/10.1017/S1473550404001715

van der Meer, J. (2006). Metabolic theories in ecology. *Trends in Ecology & Evolution*, *21*(3), 136–140. https://doi.org/10.1016/j.tree.2005.11.004

Walker, J. C. G., Hays, P. B., & Kasting, J. F. (1981). A negative feedback mechanism for the long-term stabilization of Earth's surface temperature. *Journal of Geophysical Research: Oceans*, *86*(C10), 9776–9782. https://doi.org/10.1029/JC086iC10p09776

Wierzchos, J., Davila, A. F., Artieda, O., Cámara-Gallego, B., de los Ríos, A., Nealson, K. H., Valea, S., Teresa García-González, M., & Ascaso, C. (2013). Ignimbrite as a substrate for endolithic life in the hyper-arid Atacama Desert: Implications for the search for life on Mars. *Icarus*, *224*, 334–346. https://doi.org/10.1016/j.icarus.2012.06.009

Williams, J. P., & Hallsworth, J. E. (2009). Limits of life in hostile environments: No barriers to biosphere function? *Environmental Microbiology*, *11*(12), 3292–3308. https://doi.org/10.1111/j.1462-2920.2009.02079.x

Zaks, D. P. M., Ramankutty, N., Barford, C. C., & Foley, J. A. (2007). From Miami to Madison: Investigating the relationship between climate and terrestrial net primary production. *Global Biogeochemical Cycles*, *21*(3). https://doi.org/10.1029/2006GB002705

Zsom, A. (2015). A POPULATION-BASED HABITABLE ZONE PERSPECTIVE. *The Astrophysical Journal*, *813*(1), 9. https://doi.org/10.1088/0004-637X/813/1/9

Zuluaga, J. I., Salazar, J. F., Cuartas-Restrepo, P., & Poveda, G. (2014). The Habitable Zone of Inhabited Planets. *Biogeosciences Discussions*, *11*(6), 8443–8483. https://doi.org/10.5194/bgd-11-8443-2014